\documentclass[a4paper, 8pt]{article}
\usepackage[utf8]{inputenc}
\usepackage{a4wide}

\usepackage{hyperref}

\usepackage{amsmath, amssymb, amsthm, amsfonts, mathtools, mathrsfs}

\usepackage{enumerate}

\usepackage{dsfont}

\usepackage[pdftex]{graphicx}

\usepackage{fancyhdr}

\usepackage{setspace}

\usepackage{tikz-cd}

\usepackage{float}
\restylefloat{table}
\restylefloat{figure}

\usepackage{appendix}

\usepackage[colorinlistoftodos]{todonotes}

\theoremstyle{plain}
\newtheorem{theorem}{Theorem}[section]

\newtheorem{lemma}[theorem]{Lemma}

\theoremstyle{definition}

\newtheorem{remark}[theorem]{Remark}

\begin{document}
\title{Moments of Moments of the Characteristic Polynomials of Random Orthogonal and Symplectic Matrices}
\author{Tom Claeys, Johannes Forkel, Jonathan P. Keating}

\maketitle

\begin{abstract}
\noindent Using asymptotics of Toeplitz+Hankel determinants, we establish formulae for the asymptotics of the moments of the moments of the characteristic polynomials of random orthogonal and symplectic matrices, as the matrix-size tends to infinity. Our results are analogous to those that Fahs obtained for random unitary matrices in \cite{Fahs}. A key feature of the formulae we derive is that the phase transitions in the moments of moments are seen to depend on the symmetry group in question in a significant way.
\end{abstract}

\section{Introduction}

Characteristic polynomials of large random matrices are fundamental objects in random matrix theory. Besides encoding the eigenvalues of the random matrices, they are closely connected to the theory of log-correlated fields and to Gaussian multiplicative chaos \cite{BerestyckiWebbWong, ForkelKeating, LambertOstrovskySimm, NikulaSaksmanWebb, Webb}, and their statistics show remarkable similarities with those of the Riemann zeta function and other number-theoretic $L$-functions \cite{Conreyetal, FyodorovHiaryKeating, FyodorovKeating, KeatingSnaith, KeatingSnaith2} - see \cite{BaileyKeating2} for a review. We here study moments of moments of characteristic polynomials of random matrices in the classical compact groups of orthogonal and symplectic matrices. The term {\em moments of moments} refers to the fact that we first take a moment of the characteristic polynomial with respect to the spectral variable, and then a moment with respect to the random matrix distribution. Such objects have been investigated intensively over the last years, especially because of their connection with the distribution of extreme values of characteristic polynomials and Gaussian multiplicative chaos, and because of their application to understanding the large values taken by the Riemann zeta function on the critical line and other $L$-functions \cite{FyodorovHiaryKeating, FyodorovKeating} - see \cite{BaileyKeating2} for a review. We obtain the leading order dependence of these moments of moments in the limit on the size of the random matrices when this gets large. We focus in particular on the critical points where phase transitions occur, showing that these depend in an important way on the symmetry group in question.  In the final section, we outline some applications and implications of our results.  

\subsection{Context}
Denote by $U(n)$ the group of unitary $n \times n$ matrices, by $O(n)$ the group of orthogonal $n \times n$ matrices, by $SO(n)$ the group of orthogonal $n \times n$ matrices with determinant $+1$, by $SO^-(n)$ the set of orthogonal $n \times n$ matrices with determinant $-1$, and by $Sp(2n)$ the group of $2n\times 2n$ symplectic matrices, i.e. unitary $2n\times 2n$ matrices that additionally satisfy
\begin{equation*}
UJU^T= U^TJU = J,\qquad\mbox{where}\qquad
J:= \left( \begin{array}{cc} 0 & I \\ -I & 0 \end{array} \right) \in \mathbb{R}^{2n \times 2n}.
\end{equation*}

Let $G(n) \in \left\{ U(n), \, O(n), \, SO(n), \, SO^-(n), \, Sp(2n) \right\}$. For $U \in G(n)$ let
\begin{equation*}
p_{G(n)}(\theta;U) := \text{det} \left(I-e^{-i\theta}U\right), \quad \theta \in [0,2\pi),
\end{equation*}
denote its characteristic polynomial, taken as a function on the unit circle, where all its zeroes (the phases of the eigenvalues of $U$) lie. For $\alpha > -1/2$ and $m \in \mathbb{R}$ we define the moments of moments of $p_{G(n)}(\theta;U)$ by
\begin{align}\label{def:MoM}
\begin{split}
    {\rm MoM}_{G(n)}(m,\alpha) :=& \mathbb{E}_{U \in G(n)} \left( \left( \frac{1}{2\pi} \int_0^{2\pi} |p_{G(n)}(\theta;U)|^{2\alpha} \text{d}\theta \right)^m \right),
\end{split}
\end{align}
where $\mathbb{E}_{U \in G(n)}$ denotes the expectation with respect to the normalized Haar measure on $G(n)$. We are interested in the asymptotics of ${\rm MoM}_{G(n)}(m,\alpha)$ in the limit as $n \rightarrow \infty$, for fixed $\alpha$, $m$.\\

Recently there has been a good deal of attention given to ${\rm MoM}_{U(n)}(m,\alpha)$. Fyodorov, Hiary and Keating made conjectures on the large $n$ asymptotics in \cite{FyodorovHiaryKeating}, which were specified by Fyodorov and Keating in \cite{FyodorovKeating}, and which were then supported by numerical computations and generalized in \cite{FyodorovGnutzmannKeating}. 

In the case $G(n) = U(n)$ there has also been considerable interest in the maximum of the characteristic polynomial, see for example \cite{ArguinBeliusBourgade, ChhaibiMadauleNajnudel, PaquetteZeitouni}, which is connected to the moments of moments since ${\rm MoM}_{G(n)}(1/p,p/2) = \mathbb{E}_{U \in G(n)} \left( ||p_{G(n)}( \cdot;U)||_p \right)$, and since the expectation of the $L^p$-norm approximates $\mathbb{E}_{U \in G(n)}(\max_\theta |p_{G(n)}( \theta;U)| )$ for large $p$. This suggests that the asymptotics of ${\rm MoM}_{G(n)}(m,\alpha)$ can be used to motivate conjectures for the maximum of the characteristic polynomials \cite{FyodorovHiaryKeating, FyodorovKeating}. ${\rm MoM}_{G(n)}(m,\alpha)$ is also related to the $m$-th moment of the total mass of the Gaussian Multiplicative Chaos measures arising from the characteristic polynomial of random matrices from the classical compact groups, see \cite{ForkelKeating, NikulaSaksmanWebb, Webb}. 

The conjectured asymptotics of ${\rm MoM}_{U(n)}(m,\alpha)$ were proven when $m = 2$ and $\alpha > -1/4$ by Claeys and Krasovsky, as an application of their calculation of the asymptotics of Toeplitz determinants with two merging singularities via a Riemann-Hilbert analysis \cite{ClaeysKrasovsky}, and for $m,\alpha \in \mathbb{N}$ they were proven by Bailey and Keating using an approach based on exact identities for finite $n$ \cite{BaileyKeating}. Using a combinatorial approach for $m, \alpha \in \mathbb{N}$, which involved representation theory and constrained Gelfand-Tsetlin patterns, Assiotis and Keating \cite{AssiotisKeating} proved the same results as found in \cite{BaileyKeating}, but with an alternative formula for the leading order coefficient (see also \cite{Yacine}, where the results in \cite{BaileyKeating} and \cite{AssiotisKeating} were rederived using another approach). This same combinatorial approach was then used by Assiotis, Bailey and Keating to prove asymptotic formulas for $G(n) = SO(2n)$ and $G(n) = Sp(2n)$, when $m, \alpha \in \mathbb{N}$ {\color{blue}\cite{AssiotisBaileyKeating}}. For those asymptotic formulas Andrade and Best then provided an alternative proof using the methods developed in \cite{BaileyKeating}, again with an alternative formula for the leading order coefficient.  

Using the Riemann-Hilbert approach developed in \cite{ClaeysKrasovsky}, Fahs {\color{blue}\cite{Fahs}} computed the asymptotics, up to a multiplicative constant, of Toeplitz determinants with arbitrarily many merging singularities. Using this he then proved the asymptotic formula for ${\rm MoM}_{U(n)}(m,\alpha)$ for $m \in \mathbb{N}$ and general $\alpha > 0$, however without an explicit expression for the leading order coefficient:
\begin{theorem}[{\color{blue}\cite{Fahs}}]
    For $m \in \mathbb{N}$ and $\alpha > 0$, as $n \rightarrow \infty$:
    \begin{equation*}
        {\rm MoM}_{U(n)}(m,\alpha) = \begin{cases} (1+o(1)) n^{m\alpha^2} \frac{G(1 + \alpha)^{2m}\Gamma(1 - m\alpha^2)}{G(1 + 2\alpha)^m \Gamma(1 - \alpha^2)^m}, & \alpha < \frac{1}{\sqrt{m}}, \\
        e^{\mathcal{O}(1)} n^{m\alpha^2} \log n & \alpha = \frac{1}{\sqrt{m}},\\
        e^{\mathcal{O}(1)} n^{m^2\alpha^2 + 1 - m}, & \alpha > \frac{1}{\sqrt{m}}, \end{cases}
    \end{equation*}
    where $e^{\mathcal{O}(1)}$ denotes a function that is bounded and bounded away from $0$ as $n \rightarrow \infty$, and where $G(z)$ denotes the Barnes $G$-function.
\end{theorem}
Extending Fahs' Riemann-Hilbert approach and using his asymptotics of Toeplitz determinants, Claeys, Glesner, Minakov and Yang \cite{ClaeysGlesnerMinakovYang} computed asymptotics of Toeplitz+Hankel determinants with arbitrarily many merging singularities, up to a multiplicative constant. In this paper we use their result to compute the asymptotics of the moments of moments for $G(n) \in \{O(n), \, SO(n), \, SO^-(n), \, Sp(2n)\}$, for $m \in \mathbb{N}$ and $\alpha > 0$, up to the leading order coefficient.

\subsection{Statement of Results}
For $m\in\mathbb N$ and $\alpha>0$, we define
\begin{align} \label{eqn:C pm}
\begin{split}
C^\pm(m,\alpha) := \frac{G(1+\alpha)^{2m}}{G(1+2\alpha)^m} \frac{4^{-\alpha^2m^2 \pm \alpha m}}{\pi^m} \prod_{j = 0}^{m-1} \frac{\Gamma(1 - \alpha^2 - j\alpha^2) \Gamma\left( \frac{1 - \alpha^2 \pm \alpha}{2} - j\alpha^2 \right)^2}{\Gamma(1 - \alpha^2) \Gamma \left( 1 \pm \alpha - \alpha^2(m + j) \right)},
\end{split}
\end{align}
where $\Gamma$ denotes Euler's Gamma function and $G$ denotes the Barnes $G$-function, satisfying $G(z+1)=\Gamma(z)G(z)$ and $G(0)=0$, $G(1)=1$. Note that although $C^\pm(m,\alpha)$ has poles for certain values of $\alpha$, it is well defined if 
\begin{equation}\label{eq:alphasubcrit}\alpha<\min\left\{\frac{1}{\sqrt{m}}, \frac{\sqrt{8m-3}\pm 1}{4m-2}\right\}=\begin{cases}\frac{1}{\sqrt{m}}&\mbox{if $m=2$ and $\pm=+$,}\\
\frac{\sqrt{8m-3}\pm 1}{4m-2}&\mbox{otherwise,}\end{cases}\end{equation} since in this case we have
that $1-\alpha^2-j\alpha^2$ and $ \frac{1 - \alpha^2 \pm \alpha}{2} - j\alpha^2$ are positive for $j=0,\ldots, m-1$, such that the poles of $\Gamma$ are avoided.\\

We are now ready to formulate our main results.

\begin{theorem} \label{thm:MoM Sp(2n)}
Let $m \in \mathbb{N}$ and $\alpha > 0$. Then, as $n \rightarrow \infty$:
\begin{align*}
\begin{split}
    {\rm MoM}_{Sp(2n)}(m,\alpha) = \begin{cases} (1+o(1)) (2n)^{m\alpha^2} C^-(m,\alpha), & \alpha < \frac{\sqrt{8m-3} - 1}{4m-2}, \\
    e^{\mathcal{O}(1)} n^{m\alpha^2} \log n & \alpha = \frac{\sqrt{8m-3} - 1}{4m-2},\\
    e^{\mathcal{O}(1)} n^{2(m\alpha)^2 + m\alpha -m}, & \alpha > \frac{\sqrt{8m-3} - 1}{4m-2}. \end{cases}
\end{split}
\end{align*}
\end{theorem}

\begin{theorem} \label{thm:MoM O(n)}
Let $G(n) \in \{ O(n), \, SO(n), \, SO^-(n) \}$ and $\alpha > 0$. For $m \in \mathbb{N} \setminus \{2\}$, as $n \rightarrow \infty$:
\begin{align*}
\begin{split}
    {\rm MoM}_{G(n)}(m,\alpha) = \begin{cases} (1+o(1)) n^{m\alpha^2} C^+(m,\alpha), & \alpha < \frac{\sqrt{8m-3} + 1}{4m-2}, \\
    e^{\mathcal{O}(1)} n^{m\alpha^2} \log n & \alpha = \frac{\sqrt{8m-3} + 1}{4m-2},\\
    e^{\mathcal{O}(1)} n^{2(m\alpha)^2 - m\alpha -m}, & \alpha > \frac{\sqrt{8m-3} + 1}{4m-2}. \end{cases}
\end{split}
\end{align*}
Moreover, as $n \rightarrow \infty$:
\begin{align*}
    {\rm MoM}_{G(n)}(2,\alpha) = \begin{cases} (1 + o(1)) n^{2\alpha^2} C^+(2,\alpha), & \alpha < \frac{1}{\sqrt{2}}, \\
    e^{\mathcal{O}(1)} n^{2\alpha^2} \log n & \alpha = \frac{1}{\sqrt{2}},\\
    e^{\mathcal{O}(1)} n^{4\alpha^2 - 1}, & \alpha \in \left( \frac{1}{\sqrt{2}}, \frac{\sqrt{5} + 1}{4} \right), \\
    e^{\mathcal{O}(1)} n^{4\alpha^2 - 1} \log n, & \alpha = \frac{\sqrt{5} + 1}{4}, \\
    e^{\mathcal{O}(1)} n^{8\alpha^2 - 2\alpha - 2}, & \alpha > \frac{\sqrt{5} + 1}{4}. \end{cases}
\end{align*}
\end{theorem}

\begin{remark}
Since $\mathbb{E}_{U \in O(n)}(f(U)) = \mathbb{E}_{U \in SO(n)}(f(U))/2 + \mathbb{E}_{U \in SO^-(n)}(f(U))/2$ for any measurable function $f:O(n) \rightarrow \mathbb{R}$, the above result for $G(n)=O(n)$ is in fact a simple consequence of the results for $G(n)=SO(n)$ and $G(n)=SO^-(n)$, and therefore we can restrict ourselves to $G(n) \in \{SO(n), \, SO^-(n), \, Sp(2n) \}$ in what follows.
\end{remark}

\begin{remark}We will refer to the situations where $\alpha$ is small enough such that \eqref{eq:alphasubcrit} holds as the subcritical regimes or phases. For $m\neq 2$, or when both $m=2$ and $G(n)=Sp(2n)$, there is the unique critical value $\alpha=\frac{\sqrt{8m-3} + 1}{4m-2}$ for $G(n) \in \{ O(n), \, SO(n), \, SO^-(n) \}$ or $\alpha=\frac{\sqrt{8m-3} - 1}{4m-2}$ for $G(n) = Sp(2n)$, and we will refer to values of $\alpha$ larger than the critical value as the supercritical regimes. When both $m=2$ and $G(n) \in \{ O(n), \, SO(n), \, SO^-(n) \}$, then there are the two critical values $\alpha=\frac{1}{\sqrt{2}}$ and $\alpha=\frac{\sqrt{5}+1}{4}$, and we speak of the intermediate regime if $\alpha$ lies in between the two critical values, and of the supercritical regime for $\alpha>\frac{\sqrt{5}+1}{4}$.
\end{remark}

\begin{remark}For $m, \alpha \in \mathbb{N}$, our results are consistent with the results obtained by Assiotis, Bailey and Keating \cite{AssiotisBaileyKeating}, and Andrade and Best \cite{AndradeBest}. 
\end{remark}

\subsection{Proof Strategy and Outline}
For $G(n) \in \{SO(n), \, SO^-(n), \, Sp(2n) \}$, it follows from \eqref{def:MoM} and Fubini's theorem (recall that $m\in\mathbb N$) that
\begin{align} \label{eqn:Fubini}
\begin{split}
{\rm MoM}_{G(n)}(m,\alpha) =& \int_0^{2\pi} \cdots \int_0^{2\pi} \mathbb{E}_{U \in G(n)} \left( \prod_{j = 1}^m |p_{G(n)}(\theta_j; U)|^{2\alpha} \right) \frac{\text{d}\theta_1}{2\pi} \cdots \frac{\text{d}\theta_m}{2\pi} \\
=& \int_0^{\pi} \cdots \int_0^{\pi} \mathbb{E}_{U \in G(n)} \left( \prod_{j = 1}^m |p_{G(n)}(\theta_j; U)|^{2\alpha} \right) \frac{\text{d}\theta_1}{\pi} \cdots \frac{\text{d}\theta_m}{\pi}.
\end{split}
\end{align}
For the second equality above, we used the fact that
$p_{G(n)}(-\theta; U) = \overline{p_{G(n)}(\theta; U)}$, which holds since the eigenvalues of orthogonal and symplectic matrices are $\pm 1$ or appear in complex conjugate pairs. \\

For $\theta_1,\ldots,\theta_m \in (0,\pi)$ we define the symbols  
\begin{equation*}
f_m^{(\alpha)}(z) =  \prod_{j = 1}^m |z - e^{i\theta_j}|^{2\alpha} |z - e^{-i\theta_j}|^{2\alpha}.
\end{equation*}
Then by the Baik-Rains identity \cite{BaikRains} we see that the averages in the integrand in (\ref{eqn:Fubini}) can be expressed as determinants of Toeplitz+Hankel matrices:
\begin{align} \label{eqn:Baik-Rains}
\begin{split}
\mathbb{E}_{U \in SO(2n)} \left( \prod_{j = 1}^m |p_{SO(2n)}(\theta_j; U)|^{2\alpha} \right) =& \frac{1}{2} D_n^{T+H,1} \left( f_m^{(\alpha)} \right), \\
\mathbb{E}_{U \in SO^-(2n)} \left( \prod_{j = 1}^m |p_{SO^-(2n)}(\theta_j; U)|^{2\alpha} \right) =& D_{n-1}^{T+H,2} \left( f_m^{(\alpha)} \right) \prod_{j = 1}^m \left( 2 \sin \theta_j \right)^{2\alpha}, \\
\mathbb{E}_{U \in SO(2n+1)} \left( \prod_{j = 1}^m |p_{SO(2n+1)}(\theta_j; U)|^{2\alpha} \right) =&  D_n^{T+H,3} \left( f_m^{(\alpha)} \right) \prod_{j = 1}^m \left( 2 \sin \frac{\theta_j}{2} \right)^{2\alpha}, \\
\mathbb{E}_{U \in SO^-(2n+1)} \left( \prod_{j = 1}^m |p_{SO^-(2n+1)}(\theta_j; U)|^{2\alpha} \right) =&  D_n^{T+H,4} \left(f_m^{(\alpha)} \right) \prod_{j = 1}^m \left( 2 \cos \frac{\theta_j}{2} \right)^{2\alpha}, \\
\mathbb{E}_{U \in Sp(2n)} \left( \prod_{j = 1}^m |p_{Sp(2n)}(\theta_j; U)|^{2\alpha} \right) =& D_n^{T+H,2} \left( f_m^{(\alpha)} \right),
\end{split}
\end{align}
where for a function $f$ on the unit circle
\begin{align*} 
\begin{split}
D_n^{T+H,1}(f) &:= \det \left( f_{j-k} + f_{j+k} \right)_{j,k = 0}^{n-1}, \\
D_n^{T+H,2}(f) &:= \det \left( f_{j-k} - f_{j+k+2} \right)_{j,k = 0}^{n-1}, \\
D_n^{T+H,3}(f) &:= \det \left( f_{j-k} - f_{j+k+1} \right)_{j,k = 0}^{n-1}, \\
D_n^{T+H,4}(f) &:= \det \left( f_{j-k} + f_{j+k+1} \right)_{j,k = 0}^{n-1}, 
\end{split}
\end{align*}
where $f_j$ is the $j$-th Fourier coefficient:
\begin{align*}
\begin{split}
f_j &= \frac{1}{2\pi} \int_0^{2\pi} f(e^{i\theta})e^{-ij\theta}\text{d}\theta.
\end{split}
\end{align*}
On the right hand side of \eqref{eqn:Baik-Rains}, the Toeplitz+Hankel determinants account for the contribution of the complex conjugate pairs of eigenvalues of $U$, while the extra factors are contributions
from the fixed eigenvalues at $\pm 1$. 

Uniform asymptotics of Toeplitz+Hankel determinants, or equivalently of averages of multiplicative statistics over the free eigenvalues in the ensemble, including the case when singularities are allowed to merge, were computed in \cite[Theorem 2.2]{ClaeysGlesnerMinakovYang}, up to an $e^{\mathcal{O}(1)}$ factor. These results applied to our symbols $f_m^{(\alpha)}$ and translated to our notations imply that uniformly over the entire region $0 < \theta_1 < \cdots < \theta_m < \pi$, as $n \rightarrow \infty$,
\begin{align} \label{eqn:uniform asymptotics}
\begin{split}
D_n^{T+H,1}(f_m^{(\alpha)}) =& e^{O(1)} n^{m\alpha^2} F_n(\theta_1,\ldots, \theta_m)\prod_{j = 1}^m  \left( 2\sin \theta_j + \frac{1}{n} \right)^{- \alpha^2 + \alpha},\\
D_n^{T+H,2}(f_m^{(\alpha)}) =& e^{O(1)} n^{m\alpha^2} F_n(\theta_1,\ldots, \theta_m)\prod_{j = 1}^m \left( 2\sin \theta_j + \frac{1}{n} \right)^{- \alpha^2 - \alpha} ,\\
D_n^{T+H,3}(f_m^{(\alpha)}) =& e^{O(1)} n^{m\alpha^2} F_n(\theta_1,\ldots, \theta_m)\prod_{j = 1}^m \left( 2\sin \frac{\theta_j}{2} + \frac{1}{n} \right)^{- \alpha^2 - \alpha} \left( 2\cos \frac{\theta_j}{2} + \frac{1}{n} \right)^{- \alpha^2 + \alpha} ,\\
D_n^{T+H,4}(f_m^{(\alpha)}) =& e^{O(1)} n^{m\alpha^2} F_n(\theta_1,\ldots, \theta_m)\prod_{j = 1}^m \left( 2\sin \frac{\theta_j}{2} + \frac{1}{n} \right)^{- \alpha^2 + \alpha} \left( 2\cos \frac{\theta_j}{2} + \frac{1}{n} \right)^{- \alpha^2 - \alpha} ,
\end{split}
\end{align}
where 
\begin{align*}
F_n(\theta_1,\ldots, \theta_m) = \prod_{1 \leq j < k \leq m } \left( 2\sin \left| \frac{\theta_j - \theta_k}{2} \right| + \frac{1}{n} \right)^{-2\alpha^2} \left( 2\sin \left| \frac{\theta_j +\theta_k}{2} \right| + \frac{1}{n} \right)^{-2\alpha^2}. 
\end{align*}

Let $H(n) \in \left\{ SO(2n), \, SO^-(2n), \, SO(2n+1), \, SO^-(2n+1), \, Sp(2n) \right\}$. Combining (\ref{eqn:Fubini}), (\ref{eqn:Baik-Rains}), and (\ref{eqn:uniform asymptotics}), we see that
\begin{align} \label{eqn:MoM I}
\begin{split}
    {\rm MoM}_{H(n)}(m,\alpha) =& e^{\mathcal{O}(1)} n^{m\alpha^2} I_{H(n)}(\alpha, (0,\pi)^m), 
\end{split}
\end{align}
where for a measurable subset $R \subset (0,\pi)^m$
\begin{align} \label{eqn:I}
I_{SO(2n)}(\alpha,R) :=& \int_R F_n(\theta_1,\ldots, \theta_m) \prod_{j = 1}^m \left( 2\sin \theta_j + \frac{1}{n} \right)^{- \alpha^2 + \alpha} \frac{\text{d}\theta_1}{\pi} \cdots \frac{\text{d}\theta_m}{\pi}, \nonumber \\
I_{SO^-(2n)}(\alpha,R) :=& \int_R F_n(\theta_1,\ldots, \theta_m) \prod_{j = 1}^m \left( 2\sin \theta_j + \frac{1}{n} \right)^{- \alpha^2 - \alpha} \left( 2\sin \theta_j \right)^{2\alpha} \frac{\text{d}\theta_1}{\pi} \cdots \frac{\text{d}\theta_m}{\pi}, \nonumber \\
I_{SO(2n+1)}(\alpha,R) :=& \int_R F_n(\theta_1,\ldots, \theta_m) \\
&\times \prod_{j = 1}^m \left( 2\sin \frac{\theta_j}{2} + \frac{1}{n} \right)^{- \alpha^2 - \alpha} \left( 2\cos \frac{\theta_j}{2} + \frac{1}{n} \right)^{- \alpha^2 + \alpha} \left( 2\sin \frac{\theta_j}{2} \right)^{2\alpha} \frac{\text{d}\theta_1}{\pi} \cdots \frac{\text{d}\theta_m}{\pi}, \nonumber \\
I_{SO^-(2n+1)}(\alpha,R) :=& \int_R F_n(\theta_1,\ldots, \theta_m) \nonumber \\
&\times \prod_{j = 1}^m \left( 2\sin \frac{\theta_j}{2} + \frac{1}{n} \right)^{- \alpha^2 + \alpha} \left( 2\cos \frac{\theta_j}{2} + \frac{1}{n} \right)^{- \alpha^2 - \alpha} \left( 2\cos \frac{\theta_j}{2} \right)^{2\alpha} \frac{\text{d}\theta_1}{\pi} \cdots \frac{\text{d}\theta_m}{\pi}, \nonumber \\
I_{Sp(2n)}(\alpha,R) :=& \int_R F_n(\theta_1,\ldots, \theta_m) \prod_{j = 1}^m \left( 2\sin \theta_j + \frac{1}{n} \right)^{- \alpha^2 - \alpha} \frac{\text{d}\theta_1}{\pi} \cdots \frac{\text{d}\theta_m}{\pi}. \nonumber
\end{align}
For the proofs of the subcritical regimes in Theorem \ref{thm:MoM Sp(2n)} and Theorem \ref{thm:MoM O(n)}, we will show in Section \ref{section:subcritical} that for $R = (0,\pi)^m$ the above integrals converge as $n\to\infty$ to Selberg-type integrals which can be evaluated explicitly.

In the critical, intermediate, and supercritical regimes the integrals $I_{H(n)}(\alpha,(0,\pi)^m)$ diverge, and we need to prove optimal lower and upper bounds for them, up to an $e^{\mathcal{O}(1)}$ term. \\

To obtain lower bounds we use the inequalities
\begin{align*}
    \left( 2\sin \left| \frac{\theta_j \pm \theta_k}{2} \right| + \frac{1}{n} \right)^{-2\alpha^2} \geq & \left(\frac{n}{3}\right)^{2\alpha^2}, \\ 
    \left( 2\sin \frac{\theta_j}{2} + \frac{1}{n} \right)^{-\alpha^2 \pm \alpha} \geq \left( 2\sin \theta_j + \frac{1}{n} \right)^{-\alpha^2 \pm \alpha} \geq & \left(\frac{n}{3}\right)^{\alpha^2 \mp \alpha}, \quad \text{for } -\alpha^2 \pm  \alpha < 0, \\
    \left( 2\sin \theta_j + \frac{1}{n} \right)^{-\alpha^2 \pm \alpha} \geq \left( 2\sin \frac{\theta_j}{2} + \frac{1}{n} \right)^{-\alpha^2 \pm \alpha} \geq & n^{\alpha^2 \mp \alpha}, \quad \text{for } -\alpha^2 \pm \alpha \geq 0,
\end{align*}
valid for $0<\theta_j,\theta_k < 1/n$, and the fact that
\begin{align*}
    \int_0^{1/n} (2\sin \theta_j)^{2\alpha} \text{d}\theta_j \geq \int_0^{1/n} \left( 2\sin \frac{\theta_j}{2} \right)^{2\alpha} \text{d}\theta_j \geq (2/\pi)^{2\alpha} \int_{0}^{1/n} \theta_j^{2\alpha} \text{d}\theta_j = \frac{(2/\pi)^{2\alpha}}{2\alpha + 1} n^{-2\alpha - 1},
\end{align*}
to obtain the inequalities
\begin{align} \label{eqn:supercritical lower bound}
I_{H(n)}(\alpha,(0,\pi)^m)\geq 
I_{H(n)}(\alpha,(0,1/n)^m)\geq 
cn^{2m(m-1)\alpha^2 - m(1 -\alpha^2 \pm \alpha)},
\end{align}
with $\pm = +$ for $H(n)$ one of the orthogonal ensembles, and $\pm = -$ if $H(n)=Sp(2n)$.
Together with (\ref{eqn:MoM I}) this provides us with the required lower bounds in the supercritical phases in Theorems \ref{thm:MoM Sp(2n)} and \ref{thm:MoM O(n)}, except when both $m=2$ and $H(n)\in\left\{ SO(2n), \, SO^-(2n), \, SO(2n+1), \, SO^-(2n+1)\right\}$. \\

Observe also that the lower bound diverges as $n \to \infty$ if and only if 
\begin{align*} 
2m(m-1)\alpha^2 - m(1 -\alpha^2 \pm \alpha) > 0 \iff \alpha > \frac{\sqrt{8m-3} \pm 1}{4m-2}.
\end{align*}

To prove the lower bound $c \log n$ for $I_{H(n)}(\alpha,(0,\pi)^m)$ in the critical phase $\alpha = \frac{\sqrt{8m-3} \pm 1}{4m-2}$, we define the sets 
\begin{align*}
    B_n(\ell) := (0,\ell/n)^m\setminus (0,(\ell-1)/n)^m, \quad \ell = 2,\ldots,n.
\end{align*}
On $B_n(\ell)$ it holds that
\begin{align*}
\begin{split}
    \left( 2 \sin \left|\frac{\theta_j \pm \theta_k}{2}\right| + \frac{1}{n} \right)^{-2\alpha^2} \geq& \left( \frac{2\ell+1}{n} \right)^{-2\alpha^2}, \quad 1 \leq j < k \leq m,
\end{split}
\end{align*}
and when $-\alpha^2 \pm \alpha \leq 0$, then additionally
\begin{align*}
\begin{split}
    \left( 2\sin \frac{\theta_j}{2} + \frac{1}{n} \right)^{-\alpha^2 \pm \alpha} \geq &\left( 2 \sin \theta_j + \frac{1}{n} \right)^{-\alpha^2 \pm \alpha} \geq \left( \frac{2\ell+1}{n} \right)^{-\alpha^2 \pm \alpha}, \quad 1 \leq j \leq m.
\end{split}
\end{align*}
Further we see that for all $\alpha > 0$
\begin{align*}
\begin{split}
    &\int_{B_n(\ell)} \prod_{j = 1}^m (2 \sin \theta_j)^{2\alpha} \text{d}\theta_j \geq \int_{B_n(\ell)} \prod_{j = 1}^m \left( 2 \sin \frac{\theta_j}{2} \right)^{2\alpha} \text{d}\theta_j \geq c \int_{B_n(\ell)} \prod_{j = 1}^m \theta_j^{2\alpha} \text{d}\theta_j  \\
    =& c'n^{-2m\alpha - m} \left( \ell^{2m\alpha + m} - (\ell-1)^{2m\alpha + m} \right) \geq c'' n^{-2m\alpha - m} \ell^{2m\alpha + m - 1},
\end{split}
\end{align*}
and that for $-\alpha^2 + \alpha > 0$
\begin{align*}
\begin{split}    
    &\int_{B_n(\ell)} \prod_{j = 1}^m \left(2 \sin \theta_j + \frac{1}{n} \right)^{-\alpha^2 + \alpha} \text{d}\theta_j \geq \int_{B_n(\ell)} \prod_{j = 1}^m \left(2 \sin \frac{\theta_j}{2} + \frac{1}{n} \right)^{-\alpha^2 + \alpha} \text{d}\theta_j \geq c \int_{B_n(\ell)} \prod_{j = 1}^m \theta_j^{-\alpha^2 + \alpha} \text{d}\theta_j \\
    =& c'n^{m\alpha^2 - m\alpha - m} \left( \ell^{-m\alpha^2 + m\alpha + m}  - (\ell-1)^{-m\alpha^2 + m\alpha + m}\right) \geq c''n^{m\alpha^2 - m\alpha - m} \ell^{-m\alpha^2 + m\alpha + m - 1}.
\end{split}
\end{align*}
Thus, since $B_n(\ell)$, $\ell = 2,\ldots,n$, are disjoint, and since 
\begin{equation*}
\alpha = \frac{\sqrt{8m-3} - 1}{4m-2} \iff - 2m(m-1)\alpha^2 - m\alpha^2 \pm m\alpha + m = 0, 
\end{equation*} 
there exist constants $c_1, c_2>0$, independent of $n$, such that
\begin{align*}
\begin{split}
    I_{H(n)} (\alpha, (0,\pi)^m) \geq c_1\sum_{\ell = 2}^{n} \left( \ell/n \right)^{-2m(m-1)\alpha^2 -m\alpha^2 \pm m\alpha + m} \ell^{-1} = c_1 \sum_{\ell = 2}^n \ell^{-1} > c_1 \log n - c_2.  
\end{split}
\end{align*}
This provides a sharp lower bound in the critical case, except when both $m=2$ and $H(n)$ is equal to one of the orthogonal ensembles.

The upper bounds in the critical and supercritical phases of Theorem \ref{thm:MoM Sp(2n)}, and Theorem \ref{thm:MoM O(n)} in the case $m \neq 2$, are more involved: they follow from (\ref{eqn:MoM I}) and the following lemma, which will be proven in Section \ref{section:critical and supercritical}.

\begin{lemma} \label{lemma:I}
Let $\alpha > 0$. As $n \rightarrow \infty$, we have the following estimates for $m \in \mathbb{N} \setminus \{2\}$ when $H(n) \in \left\{ SO(2n), \, SO^-(2n), \, SO(2n+1), \, SO^-(2n+1), \, Sp(2n) \right\}$, and for $m \in \mathbb{N}$ when $H(n)=Sp(2n)$,
    \begin{equation*}
        I_{H(n)}(\alpha,(0,\pi)^m) = \begin{cases} \mathcal{O}(\log n) & \alpha = \frac{\sqrt{8m-3} \pm 1}{4m-2}, \\
        \mathcal{O}\left(n^{2m(m-1)\alpha^2 - m(1 -\alpha^2 \pm \alpha)} \right) & \alpha > \frac{\sqrt{8m-3} \pm 1}{4m-2},
        \end{cases}
    \end{equation*}
    with $\pm = -$ if $H(n)=Sp(2n)$ and $\pm = +$ otherwise.
\end{lemma}

\begin{remark}
    In the subcritical phase, i.e. when $\alpha < \frac{\sqrt{8m-3} \pm 1}{4m-2}$, the integral $I_{H(n)}(\alpha,(0,\pi)^m)$ converges to the (finite) Selberg-type integral $I_\infty^\pm(\alpha,(0,\pi)^m)$, defined in (\ref{eqn:I infty}) below. 
\end{remark}

It remains to consider the case where both $H(n)\in \left\{ SO(2n), \, SO^-(2n), \, SO(2n+1), \, SO^-(2n+1)\right\}$ and $m = 2$. Then there are two additional phases. By integrating over
\begin{align*}
    \left\{ (\theta_1,\ldots,\theta_m) \in (0,\pi)^m: \left|\theta_1 - \frac{\pi}{2}\right| < \frac{\pi}{4}, \, \, \max_{1 \leq j < k \leq m} |\theta_j - \theta_k| < \frac{1}{n} \right\}, 
\end{align*}
we obtain the lower bound
\begin{align} \label{eqn:m = 2 bulk 1}
 I_{H(n)}(\alpha,(0,\pi)^m)&\geq    cn^{m(m-1)\alpha^2 + 1 - m} , 
\end{align}
which diverges if and only if 
\begin{align*}
    \alpha^2 > 1/m \text{ and } m > 1 \iff & \alpha > \frac{1}{\sqrt{m}} \text{ and } m > 1.
\end{align*}
Also, exactly as in Section 2.1.3 in \cite{Fahs}, one can show that for $\alpha = \frac{1}{\sqrt{m}}$ there exist constants $c_3,c_4$ such that
\begin{align} \label{eqn:m = 2 bulk 2}
   I_{H(n)}(\alpha,(0,\pi)^m)\geq c_3\log n - c_4.
\end{align}
For $m \geq 2$ in the case $\pm = -$, and $m \geq 3$ in the case $\pm = +$, it holds that 
\begin{align*}
    \frac{1}{\sqrt{m}} >& \, \, \frac{\sqrt{8m-3} \pm 1}{4m-2}, \\
    m(m-1)\alpha^2 + 1 - m <& \, \, 2m(m-1)\alpha^2 - m(1 -\alpha^2 \pm \alpha), \quad \forall \, \, \alpha \geq \frac{1}{\sqrt{m}},
\end{align*}
which implies that in those cases the lower bounds (\ref{eqn:m = 2 bulk 1}) and (\ref{eqn:m = 2 bulk 2}) are less sharp than the previously obtained ones in (\ref{eqn:supercritical lower bound}) and can thus be ignored. \\

However, when $m = 2$ in the orthogonal cases, it holds that $\frac{1}{\sqrt{m}} < \, \, \frac{\sqrt{8m-3} \pm 1}{4m-2}$, and the lower bounds (\ref{eqn:m = 2 bulk 1}) and (\ref{eqn:m = 2 bulk 2}) are optimal for $\alpha = \frac{1}{\sqrt{2}}$ and $\frac{1}{\sqrt{2}} < \alpha < \frac{\sqrt{5}+1}{4}$, respectively. For $\alpha = \frac{\sqrt{5}+1}{4}$, when both the lower bounds (\ref{eqn:supercritical lower bound}) and (\ref{eqn:m = 2 bulk 1}) diverge with the same power, it turns out that an extra $\log n$ term appears. The following lemma states the different phases of the asymptotics of $I_{H(n)}(\alpha,(0,\pi)^2)$ for $H(n) \in \{ SO(2n), \, SO(2n+1), \, SO^-(2n+1), \, SO^-(2n+1) \}$. It will be proven in Section \ref{section:m = 2}, and together with (\ref{eqn:MoM I}) implies Theorem \ref{thm:MoM O(n)} for $m = 2$ and the phases where $\alpha \geq \frac{1}{\sqrt{2}}$.
\begin{lemma} \label{lemma:I m = 2}
Let $\alpha \geq 1/\sqrt{2}$. For $m=2$ and $H(n)\in\{ SO(2n), \, SO^-(2n), \, SO^+(2n+1), \, SO^-(2n+1) \}$, as $n \rightarrow \infty$, it holds that
\begin{equation*}
    I_{H(n)}(\alpha, (0,\pi)^m)= \begin{cases} e^{\mathcal{O}(1)} \log n & \alpha = \frac{1}{\sqrt{2}}, \\
    e^{\mathcal{O}(1)} n^{2\alpha^2 - 1}  & \alpha \in \left( \frac{1}{\sqrt{2}}, \frac{\sqrt{5} + 1}{4} \right), \\
    e^{\mathcal{O}(1)} n^{2\alpha^2 - 1} \log n & \alpha = \frac{\sqrt{5} + 1}{4}, \\
    e^{\mathcal{O}(1)} n^{6\alpha^2 - 2\alpha - 2}  & \alpha > \frac{\sqrt{5} + 1}{4}.
    \end{cases}
\end{equation*}
\end{lemma}

\section{Proof of the subcritical phases} \label{section:subcritical}

When setting $1/n$ to zero in (\ref{eqn:I}), we obtain the integrals 
\begin{align} \label{eqn:I infty}
\begin{split}
I_\infty^\pm(\alpha,R) :=& \int_R \prod_{1 \leq j < k \leq m} \left|2\cos \theta_j - 2 \cos \theta_k \right|^{-2\alpha^2} \prod_{j = 1}^m \left|2\sin \theta_j \right|^{-\alpha^2 \pm \alpha} \frac{\text{d}\theta_1}{\pi} \cdots \frac{\text{d}\theta_m}{\pi},
\end{split}
\end{align}
with $\pm = +$ in all the orthogonal cases, and $\pm = -$ in the symplectic case. The integrals $I_\infty^\pm(\alpha,(0,\pi)^m)$ are finite in the case $m = 1$ if and only if $\alpha < \frac{\sqrt{5} \pm 1}{2}$, and in the case $m \geq 2$ they are finite if and only if $\alpha < \min \left\{ \frac{1}{\sqrt{m}}, \frac{\sqrt{8m-3} \pm 1}{4m-2} \right\}$. This follows by changing variables to $x_j = \frac{1}{2} + \frac{1}{2} \cos \theta_j$ to obtain a Selberg integral, and then using Theorem \ref{thm:Selberg} below: 
\begin{align*}
\begin{split}
& I_\infty^\pm(\alpha,(0,\pi)^m) \\
=& \frac{4^{-\alpha^2m^2 \pm \alpha m}}{\pi^m} \int_{0}^{1} \cdots \int_{0}^{1} \prod_{1 \leq j < k \leq m} |x_j - x_k|^{-2\alpha^2} \prod_{j = 1}^m (x_j(1 - x_j))^{ \frac{1 -\alpha^2 \pm \alpha}{2} - 1} \text{d}x_1 \cdots \text{d}x_m \\
=& \frac{4^{-\alpha^2m^2 \pm \alpha m}}{\pi^m} \prod_{j = 0}^{m-1} \frac{\Gamma(1 - \alpha^2 - j\alpha^2) \Gamma\left( \frac{1 - \alpha^2 \pm \alpha}{2} - j\alpha^2 \right)^2}{\Gamma(1 - \alpha^2) \Gamma \left( 1 \pm \alpha - \alpha^2(m + j) \right)}.
\end{split}
\end{align*}

\begin{theorem}[Selberg, 1944 \cite{Selberg}] \label{thm:Selberg}
We have the identity
\begin{align*}
\begin{split}
    &\int_{0}^1 \cdots \int_{0}^1 \prod_{1 \leq j < k \leq m} |x_j - x_k|^{2c} \prod_{j = 1}^m \left(1 - x_j \right)^{a - 1} x_j^{b - 1} \mathrm{d}x_1 \cdots \mathrm{d}x_m \\=& \prod_{j = 0}^{m-1} \frac{\Gamma(1 + c + jc) \Gamma\left( a + jc \right) \Gamma\left( b + jc \right)}{\Gamma(1 + c) \Gamma \left( a + b + c(m + j -1)\right)},
\end{split}
\end{align*}
where either side is finite if and only if $\Re a, \Re b > 0$, $\Re c > - \min \{1/m, \Re a/(m-1), \Re b/(m-1) \}$.
\end{theorem}    

Thus to prove the subcritical phase in Theorems \ref{thm:MoM Sp(2n)} and \ref{thm:MoM O(n)}, in view of \eqref{eqn:C pm}, we need to show that for any $\alpha$ in the subcritical phase, and for any given $\delta > 0$, there is an integer $N$ such that for all $n > N$ 
\begin{align*}
\begin{split}
\left| {\rm MoM}_{H(n)}(m,\alpha) - (2n)^{m\alpha^2} \frac{G(1+\alpha)^{2m}}{G(1+2\alpha)^m} I_\infty^\pm(\alpha,(0,\pi)^m) \right| <& \delta n^{m\alpha^2}.
\end{split}
\end{align*}
To prove this we need the following lemma, which will be proven in Section \ref{section:critical and supercritical}:
\begin{lemma} \label{lemma:I infty bound}
Let $m\in\mathbb N$, $\alpha<\min \left\{ \frac{1}{\sqrt{m}}, \frac{\sqrt{8m-3} \pm 1}{4m-2} \right\}$, and \[H(n) \in \left\{ SO(2n), \, SO^-(2n), \, SO(2n+1), \, SO^-(2n+1), \, Sp(2n) \right\}.\]
There exists an $N \in \mathbb{N}$ and a constant $C > 0$, such that for all $n > N$, and any subset $R \subset (0,\pi)^m$ which is symmetric under permutation of the variables and symmetric around $\pi/2$ in each variable, it holds that 
\begin{align*}
    I_{H(n)}(\alpha,R) \leq C I_\infty^{\pm}(\alpha,R),
\end{align*}
where $\pm=-$ if $H(n)=Sp(2n)$ and $\pm=+$ otherwise.
\end{lemma}

For $\eta > 0$ we divide $(0,\pi)^m$ into two regions $R_1(\eta)$ and $R_2(\eta)$, where $R_1(\eta)$ is the region where $ \min \{ | 2\sin \frac{\theta_j - \theta_k}{2} |, | 2\sin \frac{\theta_j + \theta_k}{2} |, |2\sin \theta_j| \} > \eta $ for all $j \neq k$, and $R_2(\eta) = (0,\pi)^m \setminus R_1(\eta)$. Then by (\ref{eqn:Baik-Rains}), (\ref{eqn:uniform asymptotics}), (\ref{eqn:I}) and Lemma \ref{lemma:I infty bound} it follows that 
\begin{align*}
\begin{split}
\int_{R_2(\eta)} \mathbb{E}_{U \in H(n)} \left( \prod_{j = 1}^m |p_{H(n)}(\theta_j; U)|^{2\alpha} \right) \frac{\text{d}\theta_1}{\pi} \cdots \frac{\text{d}\theta_m}{\pi} =& e^{\mathcal{O}(1)} n^{m\alpha^2} I_{H(n)}(\alpha,R_2(\eta)) \\
=& \mathcal O \left( n^{m\alpha^2} I^\pm_\infty(\alpha, R_2(\eta)) \right), 
\end{split}
\end{align*}
as $n \rightarrow \infty$, uniformly for $0 < \eta < \pi$, with $\pm = +$ in the orthogonal cases, and $\pm = -$ in the symplectic case. Since $I_\infty^\pm(\alpha,R_2(\eta)) \rightarrow 0$ as $\eta \rightarrow 0$, it follows that for any $\delta > 0$ we can fix an $\eta_0 > 0$ and an $N_0 \in \mathbb{N}$ such that 
\begin{align} \label{eqn:int R2}
\begin{split}
\int_{R_2(\eta)} \mathbb{E}_{U \in H(n)} \left( \prod_{j = 1}^m |p_{H(n)}(\theta_j; U)|^{2\alpha} \right) \frac{\text{d}\theta_1}{\pi} \cdots \frac{\text{d}\theta_m}{\pi} < \delta n^{m\alpha^2}/2,
\end{split}
\end{align}
for all $n \geq N_0$ and $\eta < \eta_0$. \\

We now evaluate the integral of $\mathbb{E}_{U \in H(n)} \left( \prod_{j = 1}^m |p_{H(n)}(\theta_j; U)|^{2\alpha} \right)$ over $R_1(\eta)$. When all singularities $e^{i\theta_j}$, $j = 1,...,m$, of $f_m^{(\alpha)}$ are bounded away from each other and from $\pm 1$, then Theorem 1.25 in \cite{DeiftItsKrasovsky} gives the asymptotics, including the leading order coefficient, of $D_n^{T+H,\kappa} \left( f_m^{(\alpha)} \right)$, $\kappa = 1,2,3,4$. As on $R_1(\eta)$ all singularities are bounded away from each other and from $\pm 1$, substituting those asymptotics into (\ref{eqn:Baik-Rains}) implies that
\begin{align} \label{eqn:pointwise asymptotics}
\begin{split}
&\mathbb{E}_{U \in H(n)} \left( \prod_{j = 1}^m |p_{H(n)}(\theta_j; U)|^{2\alpha} \right) = (1+o(1)) (2n)^{m\alpha^2} \frac{G(1+\alpha)^{2m}}{G(1+2\alpha)^m} \\
&\times \prod_{1 \leq j < k \leq m} \left| 2 \cos \theta_j  - 2 \cos \theta_k \right|^{-2\alpha^2} \prod_{j = 1}^m \left(2 \sin \theta_j \right)^{- \alpha^2 \pm \alpha},
\end{split}
\end{align}
as $n \rightarrow \infty$, uniformly for $(\theta_1,...,\theta_m) \in R_1(\eta)$. Combining (\ref{eqn:I infty}) and (\ref{eqn:pointwise asymptotics}) we see that
\begin{align*} 
\begin{split}
&\int_{R_1(\eta)} \mathbb{E}_{U \in H(n)} \left( \prod_{j = 1}^m |p_{H(n)}(\theta_j; U)|^{2\alpha} \right) \frac{\text{d}\theta_1}{\pi} \cdots \frac{\text{d}\theta_m}{\pi} \\
=&(1+o(1)) (2n)^{m\alpha^2} \frac{G(1+\alpha)^{2m}}{G(1+2\alpha)^m} I_\infty^\pm(\alpha,R_1(\eta)) \\
=& (1+o(1))(2n)^{m\alpha^2} \frac{G(1+\alpha)^{2m}}{G(1+2\alpha)^m} \left( I_\infty^\pm(\alpha,(0,\pi)^m) - I_\infty^\pm(\alpha,R_2(\eta)) \right),
\end{split}
\end{align*}
where the $o(1)$ term tends to zero for any fixed $\eta > 0$, as $n \rightarrow \infty$. This implies that 
\begin{align*}
\begin{split}
&\left| {\rm MoM}_{H(n)}(m,\alpha) - (2n)^{m\alpha^2} \frac{G(1+\alpha)^{2m}}{G(1+2\alpha)^m} I_\infty^\pm(\alpha,(0,\pi)^m) \right| \\
=&\left| \int_{(0,\pi)^m} \mathbb{E}_{U \in H(n)} \left( \prod_{j = 1}^m |p_{H(n)}(\theta_j; U)|^{2\alpha} \right) \frac{\text{d}\theta_1}{\pi} \cdots \frac{\text{d}\theta_m}{\pi} - (2n)^{m\alpha^2} \frac{G(1+\alpha)^{2m}}{G(1+2\alpha)^m} I_\infty^\pm(\alpha,(0,\pi)^m) \right| \\
\leq &  \left| \int_{R_1(\eta)} \mathbb{E}_{U \in H(n)} \left( \prod_{j = 1}^m |p_{H(n)}(\theta_j; U)|^{2\alpha} \right) \frac{\text{d}\theta_1}{\pi} \cdots \frac{\text{d}\theta_m}{\pi} - (2n)^{m\alpha^2} \frac{G(1+\alpha)^{2m}}{G(1+2\alpha)^m} I_\infty^\pm(\alpha,(0,\pi)^m) \right| \\
&+\int_{R_2(\eta)} \mathbb{E}_{U \in H(n)} \left( \prod_{j = 1}^m |p_{H(n)}(\theta_j; U)|^{2\alpha} \right) \frac{\text{d}\theta_1}{\pi} \cdots \frac{\text{d}\theta_m}{\pi} \\
\leq & (2n^{m\alpha^2}) \frac{G(1+\alpha)^{2m}}{G(1+2\alpha)^m} \left( o(1) I_\infty^\pm(\alpha,(0,\pi)^m) + (1 + o(1) I_\infty^\pm(\alpha,R_2(\eta)) \right) \\
&+\int_{R_2(\eta)} \mathbb{E}_{U \in H(n)} \left( \prod_{j = 1}^m |p_{H(n)}(\theta_j; U)|^{2\alpha} \right) \frac{\text{d}\theta_1}{\pi} \cdots \frac{\text{d}\theta_m}{\pi}.
\end{split}
\end{align*}
Combining this with (\ref{eqn:int R2}), the fact that $I_\infty^\pm(\alpha,R_2(\eta)) \rightarrow 0$ when $\eta \rightarrow 0$, and the fact that $o(1) \rightarrow 0$ as $n \rightarrow \infty$ for any fixed $\eta > 0$, we see that for any given $\delta > 0$ we can fix an $\eta < \eta_0$ and an $N \geq N_0$, such that 
\begin{align*}
    &\left| {\rm MoM}_{H(n)}(m,\alpha) - (2n)^{m\alpha^2} \frac{G(1+\alpha)^{2m}}{G(1+2\alpha)^m} I_\infty^\pm(\alpha,(0,\pi)^m) \right| < \delta n^{m\alpha^2},
\end{align*}
for all $n \geq N$. This finishes the proof of the subcritical phases in Theorems \ref{thm:MoM Sp(2n)} and \ref{thm:MoM O(n)}.

\section{Proof of Lemma \ref{lemma:I} and Lemma \ref{lemma:I infty bound}} \label{section:critical and supercritical}

We first see, with $\pm = +$ in all the orthogonal cases, and $\pm = -$ in the symplectic case, that
\begin{align} \label{eqn:I pm}
& I_{H(n)}(\alpha, R) \leq \int_{R} \prod_{1 \leq j < k \leq m} \left( \left|2\sin \frac{\theta_j - \theta_k}{2} \right| + \frac{1}{n} \right)^{-2\alpha^2} \left( \left| 2\sin \frac{\theta_j + \theta_k}{2} \right| + \frac{1}{n} \right)^{-2\alpha^2} \nonumber \\
&\times \prod_{j = 1}^m \left( \left| 2\sin \theta_j \right| + \frac{1}{n} \right)^{-\alpha^2 \pm \alpha} \frac{\text{d}\theta_1}{\pi} \cdots \frac{\text{d}\theta_m}{\pi} \nonumber \\
=& \int_{R} \prod_{1 \leq j < k \leq m} \left( 2|\cos \theta_j - \cos \theta_k| + \frac{1}{n} \left|2\sin \frac{\theta_j - \theta_k}{2} \right| + \frac{1}{n} \left| 2\sin \frac{\theta_j + \theta_k}{2} \right| + \frac{1}{n^2} \right)^{-2\alpha^2} \\
&\times \prod_{j = 1}^m \left( \left| 2\sin \theta_j \right| + \frac{1}{n} \right)^{-\alpha^2 \pm \alpha} \frac{\text{d}\theta_1}{\pi} \cdots \frac{\text{d}\theta_m}{\pi} \nonumber \\
\leq & C \int_{R} \prod_{1 \leq j < k \leq m} \left( |\cos \theta_j - \cos \theta_k| + \frac{1}{n^2} \right)^{-2\alpha^2} \prod_{j = 1}^m \left( |\sin \theta_j| + \frac{1}{n} \right)^{- \alpha^2 \pm \alpha} \text{d}\theta_1 \cdots \text{d}\theta_m, \nonumber
\end{align}
for a constant $C$ which is independent of $n$. Making the variable transformation $\cos \theta_j = t_j$ it follows that
\begin{align} \label{eqn:I hat}
    I_{H(n)}(\alpha, (0,\pi)^m) \leq C I_n^\pm(m,\alpha),
\end{align}
where
\[
I_n^\pm(m,\alpha) 
    := \int_{(-1,1)^m} \prod_{1 \leq j < k \leq m} \left( |t_j - t_k| + \frac{1}{n^2} \right)^{-2\alpha^2} \prod_{j = 1}^m \left( \sqrt{1 - t_j^2} + \frac{1}{n} \right)^{- \alpha^2 \pm \alpha} \frac{\text{d}t_j}{\sqrt{1 - t_j^2}} .
\] 

We bound $I_n^\pm(m,\alpha)$ by the following simpler integral:
\begin{lemma} \label{lemma:I to J}
Let $\alpha>0$ and $m\in\mathbb N$. There exists $C>0$ such that for all $n \in \mathbb{N}$ it holds that
\begin{align*}
    I_n^\pm(m,\alpha) \leq C J_n^\pm(m,\alpha),
\end{align*}
where 
\begin{align*}
    J_n^\pm(m,\alpha) = \int_{[0,1)^{m}} \prod_{1 \leq j < k \leq m} \left( |t_j - t_k| + \frac{1}{n^2} \right)^{-2\alpha^2} \prod_{j = 1}^m \left( \sqrt{t_j} + \frac{1}{n} \right)^{- \alpha^2 \pm \alpha} \frac{\text{d}t_j}{\sqrt{t_j}}.
\end{align*}
\end{lemma}

\noindent \textbf{Proof:} Due to symmetry of the integrand in the $t_j$'s we see that
\begin{align*}
\begin{split}
    I_n^\pm(m, \alpha) =& \sum_{\ell = 0}^m \binom{m}{\ell} I_n^\pm(m,\alpha,\ell) ,
\end{split}
\end{align*}
where for $\ell \in \{0,\ldots,m\}$ 
\begin{multline*}
I_n^\pm(m,\alpha,\ell) \\
:= \int_{(-1,0]^\ell \times [0,1)^{m-\ell}} \prod_{1 \leq j < k \leq m} \left( |t_j - t_k| + \frac{1}{n^2} \right)^{-2\alpha^2} \prod_{j = 1}^m \left( \sqrt{1 - t_j^2} + \frac{1}{n} \right)^{- \alpha^2 \pm \alpha} \frac{\text{d}t_j}{\sqrt{1 - t_j^2}}.
\end{multline*}
By setting $s_j = - t_j$ for $1 \leq j \leq \ell$ and $s_j = t_j$ for $\ell + 1 \leq j \leq m$ we see that  
\begin{align*}
\begin{split}
    &I_n^\pm(m,\alpha,\ell) \\
    =& \int_{[0,1)^{m}} \prod_{1 \leq j < k \leq \ell} \left( |s_j - s_k| + \frac{1}{n^2} \right)^{-2\alpha^2} \prod_{1 \leq j \leq \ell < k \leq m} \left( |s_j + s_k| + \frac{1}{n^2} \right)^{-2\alpha^2} \\
    &\times \prod_{\ell + 1 \leq j < k \leq m} \left( |s_j - s_k| + \frac{1}{n^2} \right)^{-2\alpha^2} \prod_{j = 1}^m \left( \sqrt{1 - s_j^2} + \frac{1}{n} \right)^{- \alpha^2 \pm \alpha} \frac{\text{d}s_j}{\sqrt{1 - s_j^2}}  \\
    \leq & \int_{[0,1)^{m}} \prod_{1 \leq j < k \leq m} \left( |s_j - s_k| + \frac{1}{n^2} \right)^{-2\alpha^2} \prod_{j = 1}^m \left( \sqrt{1 - s_j^2} + \frac{1}{n} \right)^{- \alpha^2 \pm \alpha} \frac{\text{d}s_j}{\sqrt{1 - s_j^2}}  \\
    =& I_n^\pm(m,\alpha, 0).
\end{split}
\end{align*}
Thus 
\begin{align*} 
    I_n^\pm(m,\alpha) \leq 2^m I_n^\pm(m,\alpha,0).
\end{align*}
Since $ 1 \leq \sqrt{1 + t_j} \leq \sqrt{2}$ for $t_j \in [0,1]$, and due to symmetry, it holds that
\begin{align*}
\begin{split} 
    I_n^\pm(m,\alpha,0) &\leq  C\int_{[0,1)^{m}} \prod_{1 \leq j < k \leq m} \left( |t_j - t_k| + \frac{1}{n^2} \right)^{-2\alpha^2} \prod_{j = 1}^m \left( \sqrt{1 - t_j} + \frac{1}{n} \right)^{- \alpha^2 \pm \alpha} \frac{\text{d}t_j}{\sqrt{1 - t_j}}, \\
    &= C\int_{[0,1)^{m}} \prod_{1 \leq j < k \leq m} \left( |t_j - t_k| + \frac{1}{n^2} \right)^{-2\alpha^2} \prod_{j = 1}^m \left( \sqrt{t_j} + \frac{1}{n} \right)^{- \alpha^2 \pm \alpha} \frac{\text{d}t_j}{\sqrt{t_j}}.
\end{split}
\end{align*}
This finishes the proof. \qed \\

We now combine the factors $\left( \sqrt{t_j} + \frac{1}{n} \right)^{- \alpha^2 \pm \alpha}$ and $t_j^{-1/2}$:
\begin{lemma} \label{lemma:J to J l}
Let $\alpha>0$ and $m\in\mathbb N$. There exists $C>0$ such that for all $n \in \mathbb{N}$ it holds that
\begin{align}\label{eq:Jnpm}
     J_n^\pm(m,\alpha) \leq C \sum_{\ell = 0}^m n^{2(m-\ell)(m-\ell-1)\alpha^2 - (m-\ell)(1-\alpha^2 \pm \alpha)} J_n^\pm(m,\alpha,\ell),
\end{align}
where for $\ell = 1,\ldots,m$
\begin{align*}
\begin{split}
    J_n^\pm(m,\alpha,\ell) :=& \int_{\left[ \frac{1}{n^2},1 \right]^\ell} \prod_{1 \leq j < k \leq \ell} \left( |t_j - t_k| + \frac{1}{n^2} \right)^{-2\alpha^2} \prod_{j = 1}^\ell t_j^{\frac{-1-(4(m-\ell)+1) \alpha^2 \pm \alpha}{2}} \text{d}t_{1} \cdots \text{d}t_\ell,
\end{split}
\end{align*}
and $J_n^\pm(m,\alpha,0) := 1$.
\end{lemma}

\noindent \textbf{Proof:} We observe that
\begin{align}\label{eq:Jn1}
\begin{split}
    J_n^\pm(m,\alpha) = \sum_{\ell = 0}^m \binom{m}{l} \int_{\left[ \frac{1}{n^2},1 \right]^\ell \times \left[0,\frac{1}{n^2} \right]^{m-\ell}} \prod_{1 \leq j < k \leq m} \left( |t_j - t_k| + \frac{1}{n^2} \right)^{-2\alpha^2} \\
    \hspace{3cm} \times \prod_{j = 1}^m \left(\sqrt{t_j} + \frac{1}{n} \right)^{-\alpha^2 \pm \alpha} t_j^{-1/2} \text{d}t_j.
\end{split}
\end{align}
The integral on the right can be rewritten as 
\begin{align}\label{integral}
\begin{split}
    \int_{\left[\frac{1}{n^2}, 1\right]^\ell \times \left[0, \frac{1}{n^2}\right]^{m-\ell}} \prod_{1 \leq j < k \leq \ell} \left( |t_j - t_k| + \frac{1}{n^2} \right)^{-2\alpha^2} \prod_{1 \leq j \leq \ell < k \leq m} \left( |t_j - t_k| + \frac{1}{n^2} \right)^{-2\alpha^2} \\
    \times \prod_{\ell + 1 \leq j < k \leq m} \left( |t_j - t_k| + \frac{1}{n^2} \right)^{-2\alpha^2} \prod_{j = 1}^m \left(\sqrt{t_j}+\frac{1}{n}\right)^{-\alpha^2 \pm \alpha} t_j^{-1/2} \text{d}t_j. 
\end{split}
\end{align}
Now using the estimates
\begin{align*}
    \left( |t_j - t_k| + \frac{1}{n^2} \right)^{-2\alpha^2} \leq n^{4\alpha^2} &\text{ for } (t_j, t_k) \in \left[0,\frac{1}{n^2}\right] \times \left[0,\frac{1}{n^2}\right], \\
    \left(|t_j - t_k| + \frac{1}{n^2}\right)^{-2\alpha^2} \leq t_j^{-2\alpha^2} &\text{ for  } (t_j, t_k) \in \left[\frac{1}{n^2}, 1\right] \times \left[0,\frac{1}{n^2}\right], \\
    \left(\sqrt{t_j} + \frac{1}{n} \right)^{-\alpha^2 \pm \alpha} \leq n^{\alpha^2 \mp \alpha} &\text{ for } t_j \in \left[0,\frac{1}{n^2}\right] \text{ and } -\alpha^2 \pm \alpha \leq 0, \\ 
    \left(\sqrt{t_j} + \frac{1}{n} \right)^{-\alpha^2 \pm \alpha} \leq \left( \frac{n}{2} \right)^{\alpha^2 \mp \alpha} &\text{ for } t_j \in \left[0,\frac{1}{n^2}\right] \text{ and } -\alpha^2 \pm \alpha > 0, \\ 
    \left(\sqrt{t_j} + \frac{1}{n} \right)^{-\alpha^2 \pm \alpha} \leq t_j^{\frac{-\alpha^2 \pm \alpha}{2}} &\text{ for } t_j \in \left[\frac{1}{n^2},1\right] \text{ and } -\alpha^2 \pm \alpha \leq 0, \\ 
    \left( \sqrt{t_j} + \frac{1}{n} \right)^{-\alpha^2 \pm \alpha} \leq \left( 4 t_j \right)^{\frac{-\alpha^2 \pm \alpha}{2}} &\text{ for } t_j \in \left[\frac{1}{n^2},1\right] \text{ and } -\alpha^2 \pm \alpha > 0,
\end{align*}
and the identity $\int_0^{\frac{1}{n^2}} t^{-1/2} \text{d}t = \frac{1}{2n}$, we see that \eqref{integral} is bounded by
\begin{align*}
    C n^{2(m-\ell)(m-\ell-1)\alpha^2 - (m-\ell)(1 - \alpha^2 \pm \alpha)} \int_{\left[\frac{1}{n^2},1\right]^\ell} \prod_{1 \leq j < k \leq \ell} \left( |t_j - t_k| + \frac{1}{n^2} \right)^{-2\alpha^2} \\
    \times \prod_{j = 1}^\ell t_j^{\frac{-1-\alpha^2 \pm \alpha}{2}-2(m-\ell)\alpha^2} \text{d}t_j,
\end{align*} 
for a suitably chosen $C>0$. Substituting this in \eqref{eq:Jn1}, we obtain the result. \qed \\

Now we are able to prove Lemma \ref{lemma:I infty bound}, which is needed to complete the proof in Section \ref{section:subcritical}, of the results in the subcritical phase:\\

\noindent \textbf{Proof of Lemma \ref{lemma:I infty bound}:} We see that $J_n^{\pm}(m,\alpha,\ell) \leq J_\infty^{\pm}(m,\alpha,\ell)$, where $J_\infty^{\pm}(m,\alpha,\ell)$ denotes the integrals one obtains when setting $1/n$ to zero in the integration ranges and integrands of $J_n^{\pm}(m,\alpha,\ell)$, $\ell = 0,\ldots,m$:
\begin{align*}
    J_\infty^\pm(m,\alpha,\ell) := \int_{\left[0,1 \right]^\ell} \prod_{1 \leq j < k \leq \ell}  |t_j - t_k|^{-2\alpha^2} \prod_{j = 1}^\ell t_j^{\frac{-1-(4(m-\ell)+1) \alpha^2 \pm \alpha}{2}} \text{d}t_j.
\end{align*}
$J_\infty^{\pm}(m,\alpha,\ell)$ is a Selberg integral and is finite by Theorem \ref{thm:Selberg} for all $\alpha$ in the subcritical phase $\alpha < \min \left\{ \frac{1}{\sqrt{m}}, \frac{\sqrt{8m-3} \pm 1}{4m-2} \right\}$ and for all $\ell = 0,\ldots,m$. Moreover, in the subcritical phase, the summands in \eqref{eq:Jnpm} contain $n$ with a negative power for $\ell = 0,\ldots,m-1$ and with power zero for $\ell = m$. Thus, by (\ref{eqn:I hat}), Lemma \ref{lemma:I to J}, and Lemma \ref{lemma:J to J l}, we see that there exists a constant $C$ such that for all $n \geq N$
\begin{align*}
    I_{H(n)}(\alpha,(0,\pi)^m) \leq CJ^\pm_{\infty}(m,\alpha,m) = C2^m I^\pm_\infty(\alpha,(0,\pi)^m).
\end{align*}
We can repeat \textit{mutatis mutandis} those estimates and arguments for subsets $R \subset (0,\pi)^m$ that are symmetric under permutation of the variables and symmetric around $\pi/2$ in each variable.
Transforming $R$ appropriately, i.e. splitting up the integration range or changing variables, we then obtain that there exists a constant $C$ such that for all $n \in \mathbb{N}$
\begin{align*}
    I_{H(n)}(\alpha,R) \leq C2^m I^\pm_\infty(\alpha,R).
\end{align*}
\qed  \\

By changing variables to $x_j = t_j^{-1}n^{-2}$ in the integrals $J_n^\pm(m,\alpha,\ell)$ from Lemma \ref{lemma:J to J l} we see that 
\begin{align*} 
    J_n^\pm(m,\alpha) \leq Cn^{2m(m-1)\alpha^2 - m(1 - \alpha^2 \pm \alpha)} \sum_{\ell = 0}^m K_n^\pm(m,\alpha,\ell), 
\end{align*}
where for $\ell = 1,\ldots,m$
\begin{align} \label{eqn:K}
\begin{split}
&K_n^\pm(m,\alpha,\ell) 
:= \int_{1/n^2}^1 \cdots \int_{1/n^2}^1 \prod_{1 \leq j < k \leq \ell} \left( \left| x_j - x_k \right| + x_jx_k \right)^{-2\alpha^2} \prod_{j = 1}^\ell x_j^{((4m - 3)\alpha^2 \mp \alpha - 3)/2} \text{d}x_j,
\end{split}
\end{align}
and $K_n^\pm(m,\alpha,0) := 1$. We see, since $(|x_j - x_k| + x_jx_k)^{-2\alpha^2} \geq 1$ for $x_j,x_k \in [0,1]$, that for $\ell = 1,...,m-1$
\begin{align*}
    K_n(m,\alpha,m) \geq K_n(m,\alpha,\ell) \int_{1/n^2}^1 \cdots \int_{1/n^2}^1 \prod_{j = \ell + 1}^m x_j^{((4m - 3)\alpha^2 \mp \alpha - 3)/2} \text{d}x_{j} \geq C K_n(m,\alpha,\ell). 
\end{align*}
Thus we see that 
\begin{align} \label{eqn:J to K}
    J_n^\pm(m,\alpha) \leq Cn^{2m(m-1)\alpha^2 - m(1 - \alpha^2 \pm \alpha)} K_n^\pm(m,\alpha,m).
\end{align}
The following lemma, combined with (\ref{eqn:I pm}), (\ref{eqn:I hat}), Lemma \ref{lemma:I to J}, Lemma \ref{lemma:J to J l}, and (\ref{eqn:J to K}), will complete the proof of Lemma \ref{lemma:I}.
\begin{lemma} \label{lemma:K}
Let $m\in\mathbb N\setminus\{2\}$ with $H(n) \in \left\{ SO(2n), \, SO^-(2n), \, SO(2n+1), \, SO^-(2n+1), \, Sp(2n) \right\}$, or $m=2$ with $H(n)=Sp(2n)$.
As $n \rightarrow \infty$, with $\pm = -$ if $H(n)=Sp(2n)$ and $\pm=+$ otherwise:
\begin{align*}
\begin{split}
    K_n^\pm (m,\alpha,m) = \begin{cases} \mathcal{O}(\log n) & \alpha = \frac{\sqrt{8m-3} \pm 1}{4m-2},\\
    \mathcal{O}(1) & \alpha > \frac{\sqrt{8m-3} \pm 1}{4m-2}.
    \end{cases}
\end{split}
\end{align*}
\end{lemma}

\noindent \textbf{Proof of Lemma \ref{lemma:K}:} For $m = 1$ the proof is immediate, thus we let $m \geq 2$ for $\pm = +$ and $m \geq 3$ for $\pm = -$. For $\alpha = \frac{\sqrt{8m-3} \pm 1}{4m - 2}$ we see that
\begin{align*}
    &K_n^\pm(m,\alpha,m) \\
    \leq & m \int_{1/n^2}^1 \int_{1/n^2}^{x_m} \cdots \int_{1/n^2}^{x_m} \prod_{1 \leq j < k \leq m} \left| x_j - x_k \right|^{-2\alpha^2} \prod_{j = 1}^m x_j^{((4m - 3)\alpha^2 \mp \alpha - 3)/2} \text{d}x_1 \cdots \text{d}x_m \\
    =& m\int_{1/n^2}^1 \int_{1/(n^2x_m)}^{1} \cdots \int_{1/(n^2x_m)}^{1} x_m^{m\frac{(4m-3)\alpha^2 \mp \alpha - 3}{2}} \prod_{1 \leq j < k \leq m-1} \left| x_mt_j - x_mt_k \right|^{-2\alpha^2} \\
    &\hspace{3cm} \times \prod_{j = 1}^{m-1} |x_m - x_mt_j|^{-2\alpha^2} t_j^{((4m - 3)\alpha^2 \mp \alpha - 3)/2} \text{d}t_1 \cdots \text{d}t_{m-1} x_m^{m-1} \text{d}x_m \\
    \leq & m\int_{1/n^2}^1 x_m^{m(m-1)\alpha^2 + m\frac{\alpha^2 \mp \alpha - 1}{2} - 1} \text{d}x_m \\
    &\times \int_{0}^1 \cdots \int_{0}^1 \prod_{1 \leq j < k \leq m-1} \left| t_j - t_k \right|^{-2\alpha^2} \prod_{j = 1}^{m-1} |1 - t_j|^{-2\alpha^2} t_j^{((4m - 3)\alpha^2 \mp \alpha - 3)/2} \text{d}t_1 \cdots \text{d}t_{m-1},
\end{align*}
where we set $x_j = t_jx_m$ for $j = 2,...,m$. At the critical value the exponent in the first integral equals $-1$, thus the first integral exactly equals $2\log n$. The second integral is a Selberg integral which is finite if and only if $\alpha < 1/\sqrt{m-1}$, $-2\alpha^2 + 1 > 0$, $((4m - 3)\alpha^2 \mp \alpha - 1)/2 > 0$, $(m-2)\alpha^2 < -2\alpha^2 + 1$ and $(m-2)\alpha^2 < ((4m - 3)\alpha^2 \mp \alpha - 1)/2 > 0$. It is easy to check that all those conditions are fulfilled for $\alpha = \frac{\sqrt{8m-3} \pm 1}{4m - 2}$, which proves the lemma for the critical value. \\

For $\frac{\sqrt{8m-3} \pm 1}{4m-2} < \alpha < 1/\sqrt{m}$ we see that 
\begin{align*}
    K_n^\pm(m,\alpha,m) \leq \int_{0}^1 \cdots \int_{0}^1 \prod_{1 \leq j < k \leq m} \left| x_j - x_k \right|^{-2\alpha^2} \prod_{j = 1}^m x_j^{((4m - 3)\alpha^2 \mp \alpha - 3)/2} \text{d}x_j.
\end{align*}
The right-hand side is a Selberg integral, which by Theorem \ref{thm:Selberg} is finite exactly when $\frac{\sqrt{8m-3} \pm 1}{4m-2} < \alpha < 1/\sqrt{m}$. \\

When $\alpha \geq 1/\sqrt{m}$, then we see that for any $\epsilon > 0$ it holds that
\begin{align*}
    \prod_{1 \leq j < k \leq m} \left( \left| x_j - x_k \right| + x_jx_k \right)^{-2\alpha^2} \leq \prod_{1 \leq j < k \leq m} |x_j - x_k|^{-\frac{2}{m} + \epsilon} \prod_{j = 1}^m x_j^{(m-1)(-2\alpha^2 + \frac{2}{m} - \epsilon)},
\end{align*}
and thus 
\begin{align*}
    K_n^\pm(m,\alpha,m) \leq \int_0^1 \cdots \int_0^1 \prod_{1 \leq j < k \leq m} |x_j - x_k|^{-\frac{2}{m} + \epsilon} \prod_{j = 1}^m x_j^{2\frac{m-1}{m} - (m-1)\epsilon + \frac{\alpha^2 \mp \alpha - 1}{2} - 1} \text{d}x_j. 
\end{align*}
The right-hand side is again a Selberg integral which by Theorem \ref{thm:Selberg} is finite if and only if $\epsilon > 0$, $2\frac{m-1}{m} - (m-1)\epsilon + \frac{\alpha^2 \mp \alpha - 1}{2} > 0$, and 
\begin{align*}
    \frac{1}{m} - \frac{\epsilon}{2} <& \frac{2}{m} - \epsilon + \frac{\alpha^2 \mp \alpha - 1}{2(m-1)} \\
    \iff \frac{\epsilon}{2} <& \frac{1}{m} + \frac{\alpha^2 \mp \alpha - 1}{2(m-1)}.
\end{align*}
For $\epsilon < \frac{2}{m}$ the third condition implies the second. For $\alpha > 0$ it holds that $\alpha^2 - \alpha \geq -1/4$ and $\alpha^2 + \alpha > 0$, thus we see that (except for $\pm = +$ and $m = 2$) the third condition holds for $\epsilon$ small enough:
\begin{align*}
    \frac{1}{m} + \frac{\alpha^2 - \alpha - 1}{2(m-1)} \geq & \frac{1}{m} - \frac{5}{8(m-1)} = \frac{3m - 8}{8m(m-1)} > 0, \\
    \frac{1}{m} + \frac{\alpha^2 + \alpha - 1}{2(m-1)} > & \frac{1}{m} - \frac{1}{2(m-1)} = \frac{m - 2}{2m(m-1)} \geq 0. 
\end{align*}
This finishes the proof of the supercritical phase. \qed 

\section{Proof of Lemma \ref{lemma:I m = 2}} \label{section:m = 2}
Let $H(n) \in \{ SO(2n), \, SO^-(2n), \, SO(2n+1), \, SO^-(2n+1) \}$. We split up the integration range of $I_{H(n)}(\alpha,(0,\pi)^2)$ in \eqref{eqn:I} into $(0,\pi/2)^2$, $(0,\pi/2) \times (\pi/2, \pi)$, $(\pi/2, \pi) \times (0, \pi/2)$, and $(\pi/2, \pi)^2$. We see that (where $SO^+(n) := SO(n)$)
\begin{align} \label{eqn:ranges}
\begin{split}
    I_{SO^{\pm}(2n)}(\alpha,(0,\pi/2)^2) =& I_{SO^{\pm}(2n)}(\alpha,(\pi/2,\pi)^2),\\
    I_{SO^{\pm}(2n)}(\alpha,(0,\pi/2) \times (\pi/2, \pi)) =& I_{SO^{\pm}(2n)}(\alpha, (\pi/2,\pi) \times (0, \pi/2)),\\
    I_{SO^{\pm}(2n+1)}(\alpha,(0,\pi/2)^2) =& I_{SO^{\mp}(2n+1)}(\alpha,(\pi/2,\pi)^2),\\
    I_{SO^{\pm}(2n+1)}(\alpha,(0,\pi/2) \times (\pi/2, \pi)) =& I_{SO^{\mp}(2n+1)}(\alpha, (\pi/2,\pi) \times (0, \pi/2)).\\
\end{split}
\end{align}
Thus it suffices to prove Lemma \ref{lemma:I m = 2} for each of the eight integrals on the left-hand sides of (\ref{eqn:ranges}). We will only prove it for $I_{SO(2n)}(\alpha,(0,\pi/2)^2)$ and $I_{SO(2n)}(\alpha,(\pi/2, \pi)^2)$, as for the other six integrals in (\ref{eqn:ranges}) the proof is essentially the same. \\

We use that $2\theta/\pi \leq \sin \theta \leq \theta$ for $0 < \theta < \pi /2$ to obtain 
\begin{align} \label{eqn:theta bound}
\begin{split}
    &I_{SO(2n)}(\alpha,(0,\pi/2)^2) \\
    \leq & C\int_{0}^{\pi/2} \int_{\theta_2}^{\pi/2} \left( \theta_1^2 - \theta_2^2 + 2 \theta_1/n + \frac{1}{n^2} \right)^{-2\alpha^2} \left(\theta_1 + \frac{1}{n} \right)^{-\alpha^2 + \alpha} \left(\theta_2 + \frac{1}{n} \right)^{-\alpha^2 + \alpha} \text{d}\theta_1 \text{d}\theta_2 \\
    \leq & C' I_{SO(2n)}(\alpha,(0,\pi/2)^2).
\end{split}
\end{align} 
We see that for all $\alpha > 0$
\begin{align*}
    \left( \theta_1^2 - \theta_2^2 + 2 \theta_1/n + \frac{1}{n^2} \right)^{-2\alpha^2} \leq n^{4\alpha^2}, 
\end{align*} 
and when additionally $-\alpha^2 + \alpha + 1 < 0$ then we can bound the integral in the middle of (\ref{eqn:theta bound}) by $\mathcal{O}(n^{6\alpha^2 - 2\alpha - 2})$. Since we get a lower bound of the same power by (\ref{eqn:supercritical lower bound}) those upper and lower bounds are optimal. \\

For $-\alpha^2 + \alpha + 1 \geq 0$ we set $\theta_j = s_j/n$ and find that the integral in the middle of (\ref{eqn:theta bound}) is equal to
\[     
    Cn^{6\alpha^2 - 2 \alpha - 2} \int_0^{n\pi/2} \int_{s_2}^{n\pi/2} \left( s_1^2 - s_2^2 + 2s_1 + 1\right)^{-2\alpha^2} (s_1 + 1)^{-\alpha^2 + \alpha} (s_2 + 1)^{-\alpha^2 + \alpha} \text{d}s_1 \text{d}s_2. \]
    After another change of variables $s_1=s$, $s_2=st$, we see that this is equal to
\begin{align*}&Cn^{6\alpha^2 - 2 \alpha - 2} \int_0^{n\pi/2} \int_{0}^{1} \left( s^2(1 - t^2) + 2st + 1\right)^{-2\alpha^2} (s + 1)^{-\alpha^2 + \alpha} (st + 1)^{-\alpha^2 + \alpha} s \text{d}t \text{d}s \\
    = & Cn^{6\alpha^2 - 2 \alpha - 2} \int_0^2 \int_{0}^{1} \left( s^2(1 - t^2) + 2st + 1\right)^{-2\alpha^2} (s + 1)^{-\alpha^2 + \alpha} (st + 1)^{-\alpha^2 + \alpha} s \text{d}t \text{d}s \\
    & + Cn^{6\alpha^2 - 2 \alpha - 2} \int_2^{n\pi/2} \int_{0}^{1/2} \left( s^2(1 - t^2) + 2st + 1\right)^{-2\alpha^2} (s + 1)^{-\alpha^2 + \alpha} (st + 1)^{-\alpha^2 + \alpha} s \text{d}t \text{d}s \\
    & + Cn^{6\alpha^2 - 2 \alpha - 2} \int_2^{n\pi/2} \int_{1/2}^{1} \left( s^2(1 - t^2) + 2st + 1\right)^{-2\alpha^2} (s + 1)^{-\alpha^2 + \alpha} (st + 1)^{-\alpha^2 + \alpha} s \text{d}t \text{d}s.
\end{align*}
We then make the following observations. 
\begin{itemize}
\item The first integral on the right-hand side is finite and non-zero.
\item The second integral is bounded above and below by
    \begin{align*}
        &\int_2^{n\pi/2} s^{-5\alpha^2 + \alpha + 1} \int_0^{1/2} (st + 1)^{-\alpha^2 + \alpha} \text{d}t \text{d}s,
    \end{align*}
    multiplied by suitable constants. It holds that
    \begin{align*}
        \int_0^{1/2} (st + 1)^{-\alpha^2 + \alpha} \text{d}t \leq \begin{cases} C' s^{-\alpha^2 + \alpha}, & -\alpha^2 + \alpha + 1 < 0, \\ C' s^{-\alpha^2 + \alpha} \log s, & -\alpha^2 + \alpha + 1 = 0, \end{cases}
    \end{align*}
    and it is easy to check that these bounds are optimal for $s \geq 2$, in the sense that the left-hand side is bounded below by $c(1_{-\alpha^2+\alpha + 1 < 0} s^{-\alpha^2 + \alpha} + 1_{-\alpha^2+\alpha + 1 = 0} s^{-\alpha^2 + \alpha}\log n)$ for some $c>0$. Thus for large $n$, the second integral is bounded by 
    \begin{align*}
        &\int_2^{n\pi/2} s^{-6\alpha^2 + 2\alpha + 1} (1_{-\alpha^2 + \alpha + 1 < 0} + 1_{-\alpha^2 + \alpha + 1 = 0} \log s) \text{d}s\\ 
        =& \mathcal{O}(1) + \mathcal{O}\left( n^{-6\alpha^2 + 2\alpha + 2} \right) + 1_{- 6 \alpha^2 + 2\alpha + 2 = 0} \mathcal{O}(\log n)
    \end{align*}
    as $n\to\infty$,
    and it is straightforward to see that this bound is optimal.
\item  Since $s \geq 2$ and $t \in [1/2,1]$ in its integration range, the third integral is bounded above and below by 
    \begin{align*}
        & \int_2^{n\pi/2} s^{-2\alpha^2 + 2\alpha + 1} \int_{1/2}^{1} \left( s^2(1 - t) + st \right)^{-2\alpha^2} \text{d}t \text{d}s,
    \end{align*}
    multiplied by suitable constants. For $\alpha > 1/\sqrt{2}$ we see that
    \begin{align*}
        \int_{1/2}^{1} \left( s^2(1 - t) + st\right)^{-2\alpha^2} \text{d}t =& \left[ \left( s^2(1 - t) + st\right)^{-2\alpha^2 + 1} \frac{(s - s^2)^{-1}}{1 - 2\alpha^2} \right]_{1/2}^1 \\
        =& \frac{(s^2 - s)^{-1}}{2\alpha^2 - 1} \left( s^{-2\alpha^2 + 1} - (s^2/2 + s/2)^{-2\alpha^2 + 1} \right) \\
        \leq & C' s^{-2\alpha^2 - 1},
    \end{align*}
    while for $\alpha = \frac{1}{\sqrt{2}}$ we see that
    \begin{align*}
        \int_{1/2}^{1} \left( s^2(1 - t) + st\right)^{-2\alpha^2} \text{d}t  =& (s^2 - s)^{-1} \log \frac{s + 1}{2} \\
        \leq & C'' s^{-2} \log s,
    \end{align*}
    and we easily obtain lower bounds of the same order. Thus for large $n$, we get the following optimal bound for the third integral
    \begin{align*}
        &\int_2^{n\pi/2} s^{-4\alpha^2 + 2\alpha} (1 + 1_{2\alpha^2 = 1}\log s) \text{d} s \\
        =& \mathcal{O}(1) + \mathcal{O} \left( n^{1 - 4\alpha^2 + 2\alpha} \right) + 1_{2\alpha^2 - 1 = 0} \mathcal{O} \left( n^{1 - 4\alpha^2 + 2\alpha} \log n \right)+ 1_{1 - 4 \alpha^2 + 2\alpha = 0} \mathcal{O}(\log n).
    \end{align*} 
\end{itemize}
Further we see that for large $n$
\begin{align*}
    &I_{SO(2n)}(\alpha,(0,\pi/2) \times (\pi/2, \pi)) \\
    \leq & C\int_{0}^{\pi/2} \int_{\pi/2}^{\pi} \left( \theta_1^2 - \theta_2^2 + 2 \theta_1/n + \frac{1}{n^2} \right)^{-2\alpha^2} \left(\pi - \theta_1 + \frac{1}{n} \right)^{-\alpha^2 + \alpha} \left(\theta_2 + \frac{1}{n} \right)^{-\alpha^2 + \alpha} \text{d}\theta_1 \text{d}\theta_2 \\
    \leq & C' + C' \int_{\pi/4}^{\pi/2} \int_{\pi/2}^{3\pi/4} \left( \theta_1^2 - \theta_2^2 + 2 \theta_1/n + \frac{1}{n^2} \right)^{-2\alpha^2} \text{d}\theta_1 \text{d}\theta_2 \\
    \leq & C'' + C'' \int_{\pi/4}^{\pi/2} \int_{\pi/2}^{3\pi/4} \left( \theta_1 - \theta_2 + 1/n \right)^{-2\alpha^2} \text{d}\theta_1 \text{d}\theta_2 \\
    =& \mathcal{O}(1) + \mathcal{O}(n^{2\alpha^2 - 1}) + 1_{1 - 2 \alpha^2 = 0} \mathcal{O}(\log n),
\end{align*}
and one can easily see that this bound is optimal as well.\\

Putting all the obtained bounds together we see that
\begin{align*}
    &I_{SO(2n)}(\alpha,(0,\pi)^2) \\
    =& \mathcal{O}(1) + 1_{1 - 2 \alpha^2 = 0} \mathcal{O}(\log n) +  \mathcal{O}(n^{2\alpha^2 - 1}) + 1_{1 - 4 \alpha^2 + 2\alpha = 0} \mathcal{O}(n^{2\alpha^2 - 1} \log n) + \mathcal{O}(n^{6\alpha^2 - 2\alpha - 2}) \\
    =& \begin{cases} \mathcal{O}(1) & \alpha < \frac{1}{\sqrt{2}}, \\
        \mathcal{O}(\log n) & \alpha = \frac{1}{\sqrt{2}}, \\
        \mathcal{O}(n^{2\alpha^2 - 1})  & \alpha \in \left( \frac{1}{\sqrt{2}}, \frac{\sqrt{5} + 1}{4} \right), \\
        \mathcal{O}(n^{2\alpha^2 - 1}) \log n & \alpha = \frac{\sqrt{5} + 1}{4}, \\
        \mathcal{O}(n^{6\alpha^2 - 2\alpha - 2})  & \alpha > \frac{\sqrt{5} + 1}{4},
        \end{cases}
\end{align*}
and since those bounds are optimal in the sense that we get lower bounds of the same order, this finishes the proof.

\section{Implications}

We now outline briefly two problems to which our results have direct applications, one coming from Physics and the other from Number Theory.

\subsection{Spectral Determinants}

Many of the central questions in the theory of Quantum Chaos relate to understanding statistical properties of quantum spectra, in the semiclassical limit, in systems whose classical dynamics is chaotic \cite{BerryBak}.  The generally accepted model for the spectral statistics of generic systems relates them to the eigenvalue statistics of ensembles of random matrices.  

One of the main ways to characterize spectral statistics is through the value distribution of the spectral determinant: consider a system with quantum Hamiltonian $H$ and energy levels $E_n$; the spectral determinant may be represented formally by
\begin{equation}
\Delta(E)=\det (E-H)=\prod_n (E-E_n).
\end{equation} 
One often needs to regularise the determinant and the product -- see, for example, \cite{KeaSieb} for details -- but this does not influence the local statistical properties.  These statistical properties are then modelled by those of the characteristic polynomials of random matrices.  

Dyson \cite{Dyson} understood that the appropriate ensemble of random matrices depends on the behaviour of the quantum system in question with respect to time-reversal.  Under certain assumptions concerning the symmetries of quantum Hilbert spaces, he developed a three-fold classification of random matrix ensembles, associating each one with a universality class of quantum systems according to time-reversal symmetry.  For example, classically chaotic systems that are not time reversal symmetric are described generically by ensembles of random matrices that are invariant under unitary transformations, such as the group $U(n)$. Therefore, in such systems, the value distribution of the spectral determinants are modelled by that of the characteristic polynomials of random matrices drawn from $U(n)$ uniformly with respect to Haar measure (or by the ensemble of complex hermitian Gaussian random matrices, which is also invariant under unitary transformations).  

The value distribution of 
\begin{equation}
\log\Delta(E)=\log\det (E-H)={\rm Tr}\log (E-H),
\end{equation} 
the integral of the energy-dependent Green function, has been conjectured, for classically chaotic systems with no time-reversal-like symmetry, to satisfy a central limit theorem, when appropriately scaled, on the basis that a corresponding theorem can be proved for characteristic polynomials of random unitary matrices \cite{KeatingSnaith} and complex hermitian Gaussian random matrices \cite{CL}.  Similarly, the moments of $\Delta(E)$ are modelled by those of the  characteristic polynomials of random unitary matrices.  Recently, the extreme value statistics of $\Delta(E)$ have been investigated in the context of the characteristic polynomials of random unitary matrices \cite{Keating2022}.  In particular, the moments of moments of $\Delta(E)$, defined by
\begin{equation}
{\rm MoM}_{H, X}(m,\alpha):=\frac{1}{X}
\int_0^X\left(\int_x^{x+1} |\Delta(E)|^{2\alpha}\text{d}E\right)^m\text{d}x,
\end{equation}
are modelled in relation to the moments of moments associated with the unitary group $U(n)$.  Hence the moments of moments in this case are expected to exhibit phase transitions akin to these described by Theorem 1.1.   

Altland and Zirnbauer \cite{AZ}  made the important observation that if one allows a wider class of Hilbert spaces than Dyson considered, for example including Hilbert spaces associated with superconducting systems in which there is particle-hole symmetry, or systems with particle-antiparticle symmetries described by the Dirac equation, then Dyson's threefold classification extends to a 10-fold classification which includes not just systems whose spectral statistics are modelled by the random matrices drawn from the unitary group, but as well systems whose spectral statistics are modelled by random matrices drawn from the orthogonal group or by the symplectic group.  In the latter cases the moments of moments need to be modelled using the results we derive here, stated in Theorem 1.2 and Theorem 1.3.  In particular, the moments of moments in these cases will exhibit different phase transitions to those described by Theorem 1.1; they will instead have the phase transition structures we establish here. 

\subsection{Number-theoretic $L$-functions}

Many of the central problems in the theory of prime numbers relate to understanding the value distribution on the critical line $\text{Re}s=1/2$ of the Riemann zeta function, defined by a Dirichlet series, or equivalently an Euler product over primes $p$,
\begin{equation}
\zeta(s):=\sum_{n=1}^\infty\frac{1}{n^s}=\prod_p\left(1-\frac{1}{p^s}\right)^{-1}
\end{equation}
for $\text{Re}s>1$, and then by analytic continuation to the rest of the complex plane.   This value distribution at a height $t$ is modelled by the characteristic polynomials of random unitary matrices of dimension approximately $\log \frac{t}{2\pi}$ \cite{KeatingSnaith}.  For example, the moments of $\log\zeta (1/2+\text{i}t)$ and of $|\zeta (1/2+\text{i}t)|$ are modelled closely by the corresponding moments of the characteristic polynomials of random unitary matrices of this size.  

Recently, there has been a good deal of interest in modelling the extreme value statistics of the zeta function on its critical line using this connection to random matrices \cite{FyodorovHiaryKeating,    
FyodorovKeating, BaileyKeating2}.  For example, there are precise conjectures for the moments of moments of the zeta function, defined by 
\begin{equation}
{\rm MoM}_{\zeta, T}(m,\alpha):=\frac{1}{T}
\int_0^T\left(\int_x^{x+1} |\zeta (1/2+\text{i}t)|^{2\alpha}\text{d}t\right)^m\text{d}x,
\end{equation}
based on Theorem 1.1 \cite{BaileyKeating3}.  These imply the moments of moments of the zeta function have the same structure of phase transitions as those described in that theorem. 

The Riemann zeta-function is one of a more general class of what are known as $L$-functions. The other $L$-functions share similar analytic properties: they all have a Dirichlet series, an Euler product, and a functional equation, like that satisfied by the zeta function.  In each case they have a critical line on which their value distribution is of considerable importance.  

It was suggested by Katz and Sarnak \cite{KatzSarnak} that $L$-functions fall into families classified by symmetry-type.  All principal $L$-functions are understood to be modelled by random unitary matrices along their critical line, but if one averages though a family, rather than along the critical line, either random unitary, orthogonal or symplectic matrices should be used, depending on the symmetry type of the family in question. For example, the set of Dirichlet $L$-functions, corresponding to twisting the Riemann zeta-function by quadratic Dirichlet characters, forms a symplectic family, and the set of quadratic twists of the $L$-function associated with a given elliptic curve form an orthogonal family. See \cite{BaileyKeating2} for further details. 

The moments of $L$-functions defined with respect to averaging through families have been explored extensively in the context of the connection to random matrices \cite{KeatingSnaith2, Conreyetal, BaileyKeating2}.  Our results here, contained in Theorems 1.2 and 1.3, are those required to extend the analysis of  \cite{BaileyKeating3} to the moments of moments of orthogonal and symplectic families of $L$-functions, and they indicate the structure of the phase transitions that we expect to see in those cases.  To be more specific, let $L_d(s)$ denote an $L$-function associated with fundamental discriminant $d$ in some family, then the results in Theorems 1.2 and 1.3 model moments of moments of the general form
\begin{equation}
{\rm MoM}_{L_d, D}(m,\alpha):=\frac{1}{D^*}
\sum_{d\le D^*}\left(\int_0^1 |L_d (1/2+\text{i}t)|^{2\alpha}\text{d}t\right)^m
\end{equation}
where $D^*$ is the number of fundamental discriminants up to $D$ (see equations (275) and (276) in \cite{BaileyKeating2} for more precise definitions).

\section*{Acknowledgements}
JF and JK were supported by ERC Advanced Grant 740900 (LogCorRM). 
TC was supported by the Fonds de la Recherche Scientifique-FNRS under EOS project O013018F.
We are grateful to Brian Conrey for very helpful discussions at the American Institute of Mathematics. 

%\newpage

\end{document}